\documentclass{elsart}
\usepackage{graphicx,wrapfig}
\usepackage{amsmath}

\begin{document}

\pagestyle{empty}

\begin{flushleft}
\large
{SAGA-HE-180-01 \hfill June 3, 2002}
\end{flushleft}
\vspace{2.4cm}
 
\begin{center}
\LARGE{\bf Studies of parton distributions} \\
{\bf at a neutrino factory} \\
\vspace{1.6cm}
\Large{S. Kumano $^*$}  \\
\vspace{0.5cm}
{Department of Physics} \\
{Saga University} \\
{Saga, 840-8502, Japan} \\
\vspace{2.4cm}
\Large{Talk given at the Third International Workshop on} \\
{Neutrino Factories based on Muon Storage Rings} \\
{(NuFACT'01)} \\
{May 24--30, 2001, Tsukuba, Japan} \\
{(talk on May 28, 2001) }  \\
\end{center}
\vspace{2.3cm}
\noindent{\rule{6.0cm}{0.1mm}} \\
\vspace{-0.3cm}
\normalsize

\noindent
{* Email: kumanos@cc.saga-u.ac.jp. 
   Information on his research is available at} \\
\noindent
{\ \, http://hs.phys.saga-u.ac.jp.}  \\

\vspace{+0.1cm}
\hfill {\large to be published in proceedings}
\vfill\eject
\setcounter{page}{1}
\pagestyle{plain}

\begin{frontmatter}
\title{Studies of parton distributions \\ at a neutrino factory}
\author{S. Kumano}
\address{Department of Physics, Saga University \\
         Honjo-1, Saga, 840-8502, Japan}
\begin{abstract}
The determination of parton distribution functions in the nucleon and nuclei
is important for obtaining precise hadron-reaction cross sections,
from which any new exotic signature could be found.
We show that a future neutrino factory could provide important information
on the parton distributions. First, a recent effort concerning
the parametrization of nuclear parton distributions is explained.
It suggests that the factory should be important for determining unknown
behavior of valence-quark distributions in nuclei at small $x$.
Second, the facility could be used for understanding nucleon spin
structure and isospin violation in the parton distributions.
\end{abstract}

\end{frontmatter}

\section{Nuclear parton distributions}
\vspace{-0.5cm}

Unpolarized parton distribution functions are now well known from
very small $x$ to relatively large $x$. However, polarized distributions
and nuclear parton distributions are not precisely determined at this stage
because of the lack of a variety of data. In the recent years,
several parametrizations of polarized distributions have been proposed. 
It was, however, unfortunate that there was no available $\chi^2$ 
analysis for nuclear distributions until recently.
An initial effort of such a nuclear parametrization is discussed in
Ref. \cite{nucl-pdf}.
The nuclear parton distributions are provided at $Q_0^2$=1 GeV$^2$ as
$f_i^A (x, Q_0^2) = w_i(x,A,Z) \, f_i (x, Q_0^2)$ with
$i$=$u_v$, $d_v$, $\bar q$, or $g$.
Here, $f_i (x, Q_0^2)$ is a distribution in the nucleon,
and $w_i(x,A,Z)$ is a weight function which takes into account
nuclear modification. It is defined with parameters
$a_i$, $b_i$, $c_i$, $d_i$, and $\beta_i$:
\begin{align}
w_i(x,A,Z) & = 1+\left( 1 - \frac{1}{A^{1/3}} \right) 
          \frac{a_i(A,Z) +b_i x+c_i x^2 +d_i x^3}{(1-x)^{\beta_i}}  .
\end{align}
Then, a $\chi^2$ analysis was performed by including electron and muon
deep inelastic experimental data. We call this analysis 
a ``cubic fit", and that without the $d_i x^3$ term is called
a ``quadratic fit". The details can be found in Ref. \cite{nucl-pdf}.

\begin{wrapfigure}{r}{0.46\textwidth}
   \vspace{-0.0cm}
     \begin{center}
     \includegraphics[width=5.5cm]{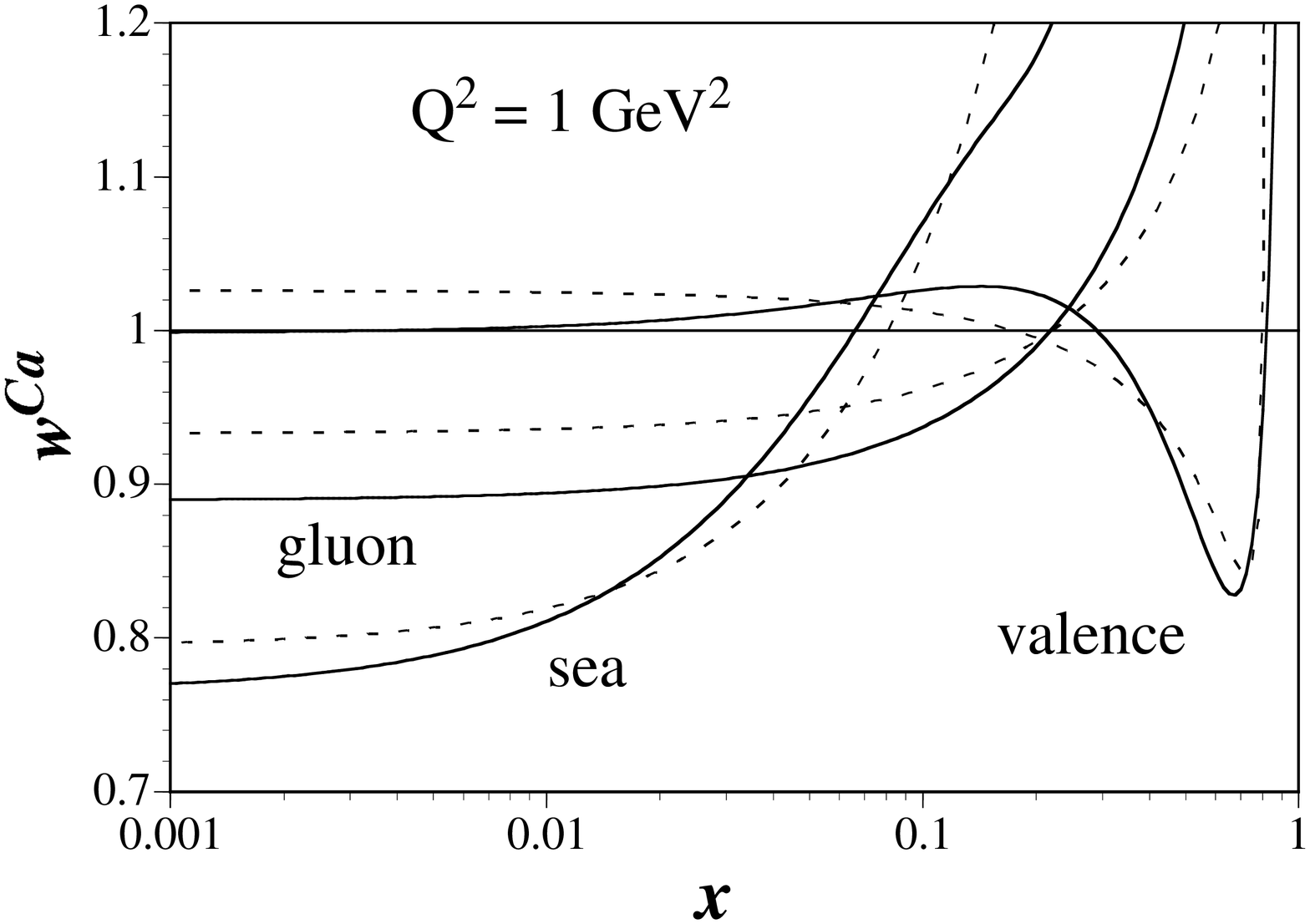}
     \end{center}
 \vspace{-0.5cm}
   \caption{\footnotesize Obtained weight functions.}
   \label{fig:wax}
\end{wrapfigure}
The obtained weight functions for the calcium nucleus
are shown in Fig.\ref{fig:wax},
where the solid curves are cubic-fit results and the dashed curves
are quadratic ones. A determination of the valence-quark distribution
at small $x$ is not possible from only the electron and muon data.
Model predictions are also confusing in the sense that
some models predict shadowing for the valence quark, and the others
predict antishadowing \cite{kkm-f3}. However, the parity-violating
structure function ($F_3$) in the neutrino interaction provides
an important clue in determining the valence distributions at small $x$:
\begin{equation}
\frac{1}{2} [ F_3^{\nu (p+n)/2} + F_3^{\bar \nu (p+n)/2} ]
= u_v +d_v +(s - \bar s)+(c - \bar c) .
\end{equation}
Because $s - \bar s$ and $c - \bar c$ are expected to be small,
neutrino-factory experiments provide accurate determination
of the valence distributions in nuclei.

There are available neutrino deep inelastic data, which are indeed
used in the nucleon parametrization. Although they play an important 
role in determining the valence-quark distribution in the ``nucleon",
the used target has been iron. Nuclear modification is not properly
taken into account in the parametrization. This fact suggests that
the valence distributions in the nucleon should be re-investigated
with the nuclear modification. If a neutrino factory becomes
available, it should be possible to take accurate deuteron data, namely
the ratios $F_3^A/F_3^D$. Then, the neutrino data can be accommodated
into the aforementioned nuclear parametrization.

\vspace{-0.7cm}
\section{Selected topics}

\vspace{-0.6cm}
\noindent
\underline{Polarized valence-quark distributions}
\vspace{-0.2cm}

We have also investigated optimum polarized parton distributions
in the nucleon \cite{aac} by analyzing polarized electron and muon
deep inelastic data. The situation is similar to the nuclear case
in the previous section in the sense that each parton distributions
cannot be well determined without accurate hadron collider data.
In the same way as $F_3$, a parity-violating structure function ($g_3$)
exists. By combining proton and neutron structure functions, we obtain
in the leading order of $\alpha_s$ \cite{lr}:
\begin{align}
g_3^{\nu p} + g_3^{\bar \nu p} 
& = - (\Delta u_v +\Delta d_v ) 
    -(\Delta s - \Delta \bar s)-(\Delta c - \Delta \bar c) ,
\nonumber  \\
g_3^{\bar \nu (p+n)/2} - g_3^{\nu (p+n)/2}
& = (\Delta s + \Delta \bar s)-(\Delta c + \Delta \bar c) .
\end{align}
These equations suggest that the neutrino facility is valuable
not only for the polarized valence-quark distributions, but also
for the antiquark distributions.

\noindent
\underline{Flavor asymmetry in light antiquark distributions}
\vspace{-0.2cm}

The $\bar u/\bar d$ asymmetry in the nucleon is now well established
by the NMC's Gottfried-sum-rule (GSR) violation and succeeding
Drell-Yan experiments. It is nonetheless important to confirm it
in a totally different process, such as the neutrino reaction.
Furthermore, it should be important to find unknown polarized
flavor asymmetry. These distributions are expressed in terms of
the neutrino structure functions as \cite{sk-pr}
\vspace{-0.4cm}
\begin{align}
\bar u - \bar d & =
    \frac{1}{4} [ F_2^{\nu p}/x - F_3^{\nu p} ]
   -\frac{1}{4} [ F_2^{\nu n}/x - F_3^{\nu n} ] ,
\nonumber  \\
\Delta \bar u - \Delta \bar d & =
    \frac{1}{2} [ g_1^{\nu p} + g_3^{\nu p} ]
   -\frac{1}{2} [ g_1^{\nu n} + g_3^{\nu n} ] .
\end{align}

\vspace{-0.5cm}
\noindent
\underline{Isospin symmetry}
\vspace{-0.2cm}

The isospin symmetry in the parton distributions is taken for granted.
In fact, it is generally believed that the isospin-violation is of
the order of a few \% ($\sim \alpha$). By combining the $F_2$ 
structure functions, we could test this ``common sense"
at the neutrino factory in connection with the GSR violation \cite{sk-pr}
\begin{align}
\int \frac{dx}{4x} \, [  F_2^{(\nu+\bar \nu) p} 
                       - F_2^{(\nu+\bar \nu) n}  ] & = 
\int dx \, [ \, 
 (\bar u + \bar d + \bar s + \bar c)_p
-(\bar u + \bar d + \bar s + \bar c)_n \, ]
\nonumber  \\ 
& = 0 \ \ \ \text{if isospin symmetry is satisfied},
\nonumber  \\ 
& \neq 0 \ \ \ \text{if the GSR violation is partly due to}
\nonumber  \\ 
& \ \ \ \ \ \ \ \ \,  \text{isospin symmetry violation}.
\end{align}

\vspace{-1.2cm}
\section{Summary}
\vspace{-0.8cm}

We learned that a future neutrino factory could provide
important information for the parton distributions in
the nucleon and nuclei. Because the distributions are essential
for calculating precise hadron cross sections, our studies are important
for finding a new signature beyond the current theoretical framework. 

\vspace{-0.6cm}
\section*{Acknowledgments}
\vspace{-0.8cm}

S.K. was supported by a Grant-in-Aid for Scientific Research
from the Japanese Ministry of Education, Culture, Sports, Science,
and Technology. He also thanks RIKEN for supporting his participation
in this conference.

\vspace{-0.5cm}


\end{document}